\documentclass{aa}  
\usepackage{lineno}
\usepackage{graphicx}
\usepackage{txfonts}
\usepackage{natbib}
\usepackage{color}
\usepackage{hyperref}
\hypersetup{colorlinks=True,citecolor=blue,allcolors=blue}
\usepackage{tabularx}
\usepackage{tabu}
\usepackage{float}
\usepackage{multirow}
\pdfoutput=1 

\begin{document} 
\nolinenumbers
   \title{Revealing the structure of the lensed quasar Q 0957+561}

   \subtitle{III. Constraints on the  size of the broad-line region}

\author{C. Fian\inst{1}, J. A. Mu\~noz\inst{1,2}, E. Mediavilla\inst{3,4}, J. Jim\'enez-Vicente\inst{5,6}, V. Motta\inst{7}, D. Chelouche\inst{8,9}, A. Wurzer\inst{10}, A. Hanslmeier\inst{10}, K. Rojas\inst{11}}
\institute{Departamento de Astronom\'{i}a y Astrof\'{i}sica, Universidad de Valencia, E-46100 Burjassot, Valencia, Spain; \email{carina.fian@uv.es} \and Observatorio Astron\'{o}mico, Universidad de Valencia, E-46980 Paterna, Valencia, Spain  \and Instituto de Astrof\'{\i}sica de Canarias, V\'{\i}a L\'actea S/N, La Laguna 38200, Tenerife, Spain \and Departamento de Astrof\'{\i}sica, Universidad de la Laguna, La Laguna 38200, Tenerife  \and Departamento de F\'{\i}sica Te\'orica y del Cosmos, Universidad de Granada, Campus de Fuentenueva, 18071 Granada, Spain \and Instituto Carlos I de F\'{\i}sica Te\'orica y Computacional, Universidad de Granada, 18071 Granada, Spain \and Instituto de F\'{\i}sica y Astronom\'{\i}a, Universidad de Valpara\'{\i}so, Avda. Gran Breta\~na 1111, Playa Ancha, Valpara\'{\i}so 2360102, Chile \and Department of Physics, Faculty of Natural Sciences, University of Haifa,
Haifa 3498838, Israel \and Haifa Research Center for Theoretical Physics and Astrophysics, University of Haifa,
Haifa 3498838, Israel \and Institute of Physics (IGAM), University of Graz, Universit{\"a}tsplatz 5, 8010, Graz, Austria \and Institute of Cosmology and Gravitation, University of Portsmouth, Burnaby Rd, Portsmouth PO1 3FX, UK}


  \abstract
  {}
   {Our aim is to examine the size, kinematics, and geometry of the broad-line region (BLR) in the double-lensed quasar \mbox{Q 0957+561} by analyzing the impact of microlensing on various rest-frame ultraviolet broad-emission lines (BELs).}
   {We explore the influence of intrinsic variability and microlensing on the C IV, C III], and Mg II emission lines through multiple spectroscopic observations taken between April 1999 and January 2017. By utilizing the line cores as a reference for no microlensing and correcting for the long time delay between the images, we estimate the sizes of the regions emitting the broad-line wings using a Bayesian approach.}
   {Our study of the microlensing amplitudes between the lensed images of the quasar Q 0957+561 reveals differing sizes of the regions emitting the three prominent BELs C IV, C III], and Mg II. The strength of the differential microlensing indicates that the high-ionization line C IV arises from a compact inner region of the BLR with a half-light radius of $R_{1/2} \gtrsim 16.0$ lt-days, which represents a lower limit on the overall size of the BLR and is comparable to the size of the region emitting the r-band continuum in this system. A somewhat larger size of $R_{1/2}\gtrsim 44$ lt-days is obtained for the semi-forbidden line \mbox{C III]}. Microlensing has a weak impact on the lower-ionization line Mg II, which is emitted from a region with a half-light radius of $R_{1/2} \gtrsim 50$ lt-days. These findings suggest that the BEL regions may have distinct geometries and kinematics, with the more extended ones being spherically symmetric, and the most compact ones being nonspherical, with motions likely confined to a plane.}
   {}

\keywords{gravitational lensing: strong -- gravitational lensing: micro -- quasars: general -- quasars: emission lines -- quasars: individual (Q 0957+561)}

\titlerunning{Broad-Line Region Size of Q 0957+561}
\authorrunning{Fian et al.} 
\maketitle

\section{Introduction}
\begin{table*}[ht]
\tabcolsep=0.85cm
\caption{Spectroscopic data.\label{data} }
\begin{tabular}{ccccc} \hline \hline \vspace*{-3mm}\\
Epoch & Date & Observed BEL & Facility & Reference\\ \hline \vspace*{-3mm} \\
1a & 15-04-1999 & C IV, C III], Mg II & HST & \citealt{Goicoechea2005}\\ 
1b & 02-06-2000 & C IV, C III], Mg II & HST & \citealt{Goicoechea2005} \\
2 & 12-01-2008 &  C IV, C III], Mg II & MMT & \citealt{Motta2012} \\
3 & 29-01-2009$^{(*)}$ & C IV, C III] & NOT & GLENDAMA\\
4 & 03/2010$^{(*)}$ & C IV, C III] & NOT & GLENDAMA\\ 
5 & 10/2010$^{(*)}$ & Mg II & LT & GLENDAMA\\ 
6 & 03/2011$^{(*)}$ & Mg II & LT & GLENDAMA\\ 
7 & 04/2011$^{(*)}$ & Mg II & LT & GLENDAMA\\ 
8 & 12/2011$^{(*)}$ & Mg II & LT & GLENDAMA\\ 
9 & 18-12-2011$^{(*)}$ & C IV, C III] & NOT & GLENDAMA\\ 
10 & 14-03-2013$^{(*)}$ & C IV, C III] & NOT & GLENDAMA\\ 
11 & 05-03-2015 & C III], Mg II & LT & GLENDAMA\\ 
12 & 19-11-2015$^{(*)}$ & C III], Mg II & LT & \citealt{GilMerino2018}\\ 
13 & 12-03-2016$^{(*)}$ & C IV, C III], Mg II & WHT & \citealt{Fian2021blr}\\
14 & 17-01-2017 & C III], Mg II & LT & \citealt{GilMerino2018}\\ \hline 
\end{tabular}\\ 

\small  \textbf{Notes.} $^{(*)}$The spectra from multiple observations conducted within a short time frame (such as several exposures taken on the same night or several observations within a month) were combined by taking their average.
\end{table*}
The twin quasar Q 0957+561 was discovered in 1979 by Dennis Walsh (\citealt{Walsh1979}) and was the first gravitational lens system to be identified. The quasar is lensed into two bright images (with a separation of $\sim$6\arcsec) by a giant elliptical lens galaxy at a redshift of $z_l=0.36$, which is part of a galaxy cluster that also contributes to the lensing (\citealt{Stockton1980,Garrett1992}). The redshift of Q 0957+561 is $z_s=1.41$, causing a significant portion of its ultraviolet (UV) emission to be observed at optical wavelengths. Multi-wavelength observations of the temporal evolution of magnification ratios in Q 0957+561 have generated a wealth of monitoring data, making it an attractive target for studying physical phenomena taking place in the lens galaxy, such as gravitational microlensing by stars and extinction by gas and dust clouds (see, e.g., \citealt{Zuo1997,Goicoechea2005,Motta2012}), as well as physical processes taking place in the background source itself, such as intrinsic variability of the quasar (see, e.g., \citealt{Lloyd1981,Miller1981,Gondhalekar1982,Planesas1999,Hutchings2003}). Optical monitoring of lensed quasars has revealed a diverse range of intrinsic flux variations, which can be used to determine accurate time delays between quasar images. In the case of \mbox{Q 0957+561}, image A leads image B by 417 days (\citealt{Shalyapin2008}).
These time delays can be used to constrain the Hubble constant (\citealt{Wong2020,Millon2020,Napier2023}) and lensing mass distributions (\citealt{Acebron2022,Fores2022}), and serve as a powerful probe of dark energy (\citealt{Wang2022,Liu2022}). Compact objects (i.e., stars) in galaxies can cause extrinsic variations in the photometric and spectroscopic observations of lensed quasars through a phenomenon known as microlensing. This effect can be used to constrain the sizes of the continuum-emitting sources surrounding the central supermassive black holes (\citealt{Fian2016,Fian2018ad,Fian2021ad,Cornachione2020,Cornachione2020b}) in order to reveal the structures  of the broad-line regions (BLRs; \citealt{Rojas2020,Hutsemekers2021,Fian2018blr,Fian2021blr}), and to estimate the masses of the black holes \citep{Mediavilla2018,Mediavilla2019,Fian2022}.\\

Our understanding of the geometry and kinematics of BLRs in quasars remains limited. To investigate the impact of microlensing on broad emission lines (BELs), it is necessary to compare spectroscopy obtained from multiple observations. Two decades of observations of Q 0957+561 have facilitated a comprehensive examination of its temporal evolution. Despite extensive photometric and spectroscopic monitoring of the quasar components A and B, the history of their magnification ratios remains a mystery. \citet{Goicoechea2005} attempted to explain the observed magnification ratios through either a dust system located between the quasar and the observer (differential extinction), or a population of microlenses in the deflector. They found that the flux ratios are consistent with both alternatives, and even a mixed scenario (extinction + microlensing) is possible. \citet{Motta2012} calculated the dust-extinction curve using BEL ratios and were able to differentiate it from microlensing. Thus, microlensing has become the favored explanation for the anomalous optical continuum ratios, supported by clear evidence of microlensing in the r-band light curves of the system (see, e.g., \citealt{Fian2021ad,Cornachione2020}). This concept was initially proposed by \citet{Chang1979} soon after the discovery of the quasar.\\

In this study, we undertake a thorough analysis of the time-variable magnification ratios in the wings of the BELs of C IV, C III, and Mg II. Our examination of gravitational microlensing and intrinsic variability involves estimating the magnitude differences between images (after correcting for the time delay), and the amplitude of variation in a single image over multiple epochs of observation. The paper is structured as follows. In Section \ref{2}, we present the spectra obtained from the literature. Section \ref{3} outlines the analysis of extrinsic and intrinsic variability in the BELs. Our method for constraining the size of the broad-line emitting regions is presented in Section \ref{4}. Finally, in Section \ref{5}, we provide conclusions based on our findings.

\begin{figure*}[ht]
\includegraphics[width=18.5cm]{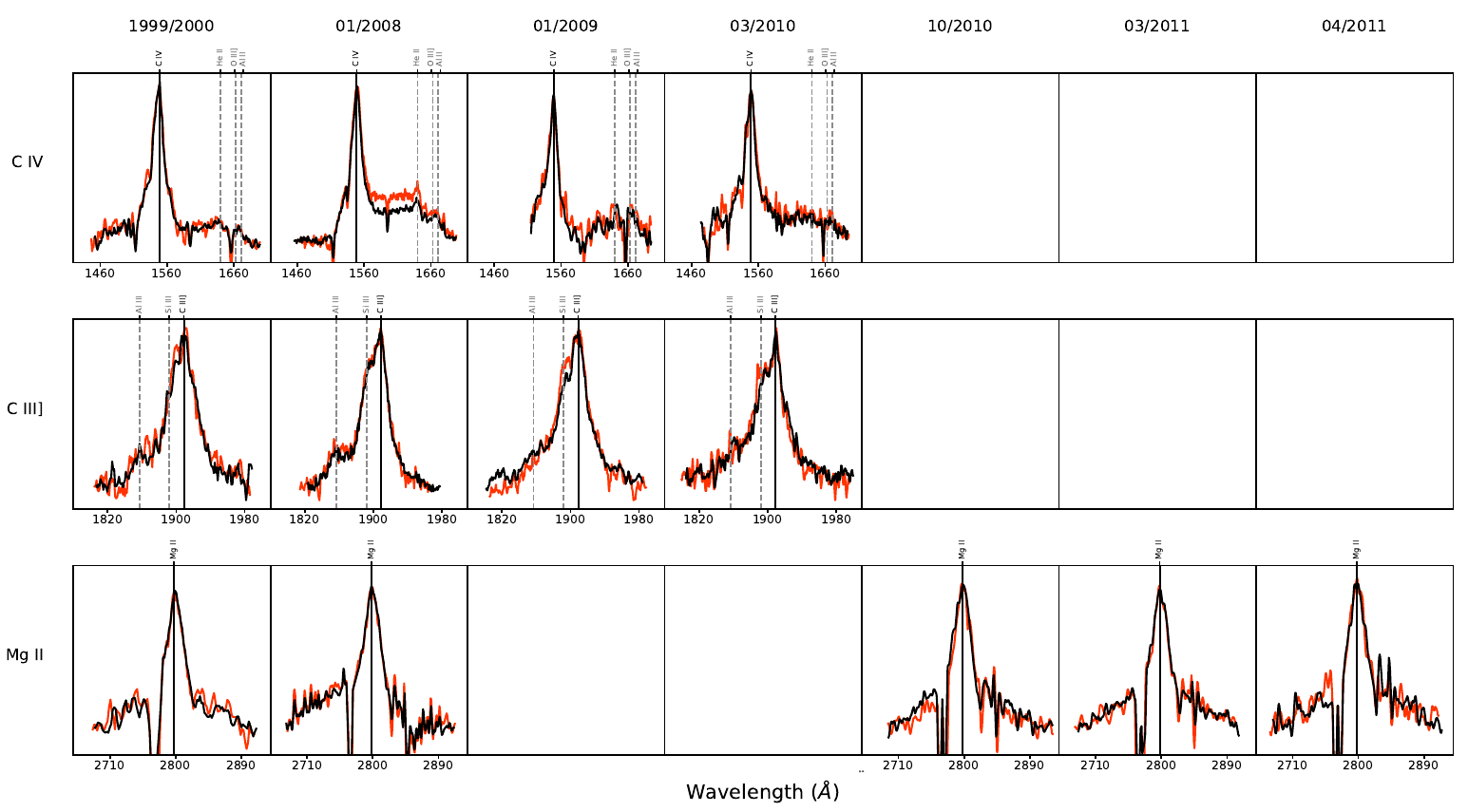}
\includegraphics[width=18.5cm]{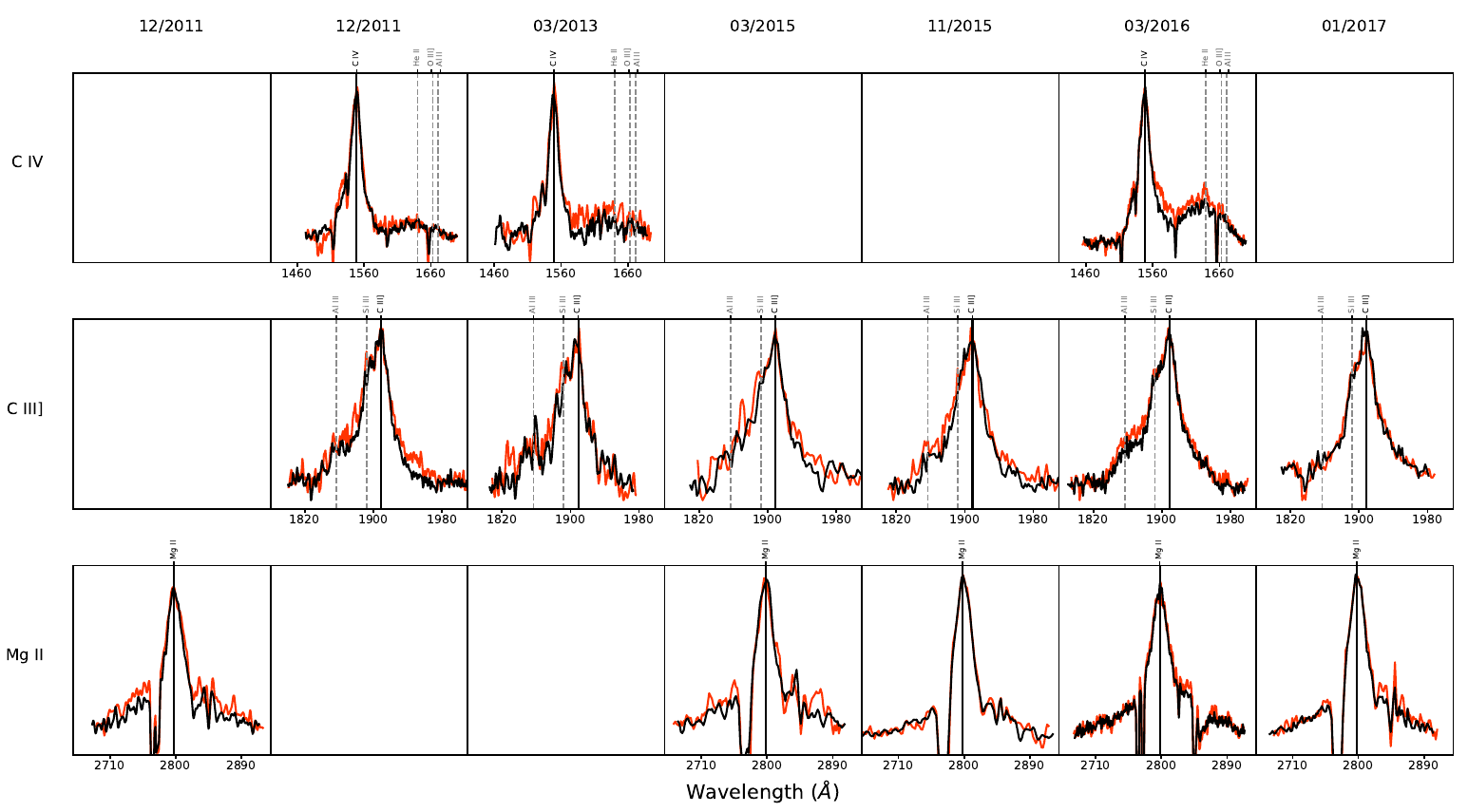}
\caption{Rest-frame emission line profiles of images A (black) and B (red) for C IV (top), C III] (middle), and Mg II (bottom) displayed in different epochs, with the continuum subtracted and the line cores matched. The leftmost panels depict the spectra of image A taken in April 1999 and image B taken 14 months later (corresponding to the time delay in this system) in June 2000.}
\label{emissionlines1}
\end{figure*}

\section{Data and observations}\label{2}
In this study, we analyzed rest-frame UV spectra of images A and B in the lensed quasar Q 0957+561. Our data set consists of 15 epochs of observation spanning a period of 18 years (1999-2017). The initial spectra of the lens system were acquired by \citet{Goicoechea2005} using the 2.4m Hubble Space Telescope (HST). \citet{Motta2012} later conducted observations with the 6.5m Multiple Mirror Telescope (MMT) in 2008. Most of the spectra were provided by the Gravitational LENses and DArk MAtter (GLENDAMA\footnote{https://grupos.unican.es/glendama/}) project of the University of Cantabria (see \citealt{GilMerino2018}). The GLENDAMA observations were carried out with the 2.0m Liverpool Telescope (LT) and the 2.6m Nordic Optical Telescope (NOT). Additionally, we obtained spectra in March 2016 using the 4.2m William Herschel Telescope (WHT) located at the Roque de los Muchachos in La Palma, Canary Islands. The data from the literature were already fully reduced. The emission lines for the quasar components A and B are displayed in Figure \ref{emissionlines1}, with observing information and references listed in Table \ref{data}.

\section{Methods}\label{3}
The estimation of microlensing signals in lensed quasars through the analysis of BELs can be complicated by the presence of intrinsic variability that is time-delayed between different images. Intrinsic variability can alter the shape of the BELs and mimic microlensing, leading to erroneous source-size estimates. To accurately measure magnitude differences between images, we select spectra that are separated in time approximately by the time delay between the images. Time delays in gravitationally lensed quasars are believed to be unique numbers that can be measured with high precision given good-quality light curves and models for the contaminating effects of gravitational microlensing.
In the absence of microlensing, the time delay for both the continuum and broad-line emitting regions should be identical. However, as demonstrated by \citet{Liao2020}, gravitational microlensing can lead to variations in the time delays measured at different wavelengths. Recent work by \citet{Tie2018} shows that gravitational microlensing produces changes in the actual time delays on the order of days. In this study, we account for this effect by incorporating an additional lag of +/- 20 days, which corresponds to the scale of the light-crossing time of the accretion disk in this system, as estimated in \citet{Fian2021ad} and \citet{Cornachione2020}. Of the available spectra, five pairs fulfill this criterion, although not all emission lines have been observed in all the selected epochs. This results in two epoch pairs (1a--1b, 3--4) covering the C IV line, three epoch pairs (1a--1b, 3--4, 12--14) including the C III] line, and three epoch pairs (1a--1b, 5--8, 12--14) containing the Mg II line. \\ 

The scale of the microlensing pattern caused by stars and compact objects in a lens galaxy (or cluster) is set by the Einstein radius, $r_E$. Significant microlensing fluctuations occur when the source size, $r_s$, is comparable to or smaller than $r_E$. The amplitude of these fluctuations will then be controlled by $r_s/r_E$, with smaller ratios leading to larger amplitudes. If the observational epochs are separated in time by more than the microlensing event timescale, also known as the Einstein radius crossing time, $t_E$, microlensing measurements are likely to be independent. This timescale is determined by the effective transverse velocity of the source, $v$, and $r_E$, with $t_E = r_E/v$ (\citealt{Paczynski1986}). \citet{Mosquera2011} report values of $r_E = 3.25 \times 10^{16} \sqrt{M/0.3M_\odot}$ cm and $t_E = 12.39$ years for the lens system Q 0957+561. The adoption of transverse peculiar velocity estimates ($\sigma_{pec}\sim640$ km s$^{-1}$ for $z_l \sim0.5$) from \citet{Mediavilla2016} results in an approximately 50\%\ reduction in $t_E$. Therefore, the criterion of independence is presumed to be met for microlensing measurements obtained from the selected pairs of observations.\\ 

As an initial step, the continuum for each image and each emission line is removed by fitting a straight line to the continuum regions adjacent to the emission line. To account for varying line widths, we use windows of varying widths to estimate the continuum, avoiding regions of known emission features. The line cores are believed to be produced by material spread over a large region (narrow-line region and outer regions of the BLR) and are therefore thought to be less susceptible to microlensing. Adopting the conclusion from \citet{Fian2018blr} that emission line cores are relatively insensitive to microlensing by solar mass objects, we establish a baseline of no microlensing by using the cores as a reference. To investigate whether the same holds true for the core of the high-ionization line C IV in Q 0957+561, we employed a method that involves fitting multiple straight lines to the continuum in the wavelength range between C IV and \mbox{C III]}, subtracting them, and subsequently normalizing the cores of the lower-ionization line C III]. We opted for this approach since \mbox{C III]} is expected to be less prone to microlensing and instrumental/calibration issues than C IV, and also to mitigate the larger extinction effects that may arise from a direct comparison with \mbox{Mg II}. Our analysis revealed that, on average, the C IV core exhibits minor variations of $(B/A)_{core} = - 0.02 \pm 0.08$ mag (68\% confidence interval), indicating that there appears to be little core variability in this system. Therefore, it is reasonable to conclude that microlensing effects are unlikely to have a significant impact on the BEL cores, which lends support to the reliability of the microlensing results presented in this work. To utilize the line cores as a reference for no microlensing, we defined the flux within a narrow interval ($\pm$ 6\AA) centered on the peak of the line and normalized the emission line cores of images A and B accordingly. Normalizing the line cores by multiplying the spectrum of image B to match the core flux of image A also effectively removes the effects of macro magnification and differential extinction (see, e.g., \citealt{Guerras2013}). After subtracting the continuum and matching the line cores, we isolate the line cores from the wings by a buffer of a few \AA ngstr\"oms to accurately assess variability in the wings. We then estimate the differential microlensing in the BELs by determining the average wing emission in velocity intervals of $\sim$5000 km s$^{-1}$ on either side of the line. In those cases in which absorption lines affect the emission line wing, a narrower integration window was chosen (blue wing of C IV) or the estimation was omitted (blue wing of Mg II) to mitigate their impact. The magnitude differences (caused by intrinsic variability and/or microlensing variability) between different observational epochs for a given image can be estimated in a similar way. The results are listed in Tables \ref{BLR_ML} and \ref{BLR_IV}.\\

The analysis of Figure \ref{emissionlines1} and Tables \ref{BLR_ML}--\ref{BLR_IV} (with special attention being paid to the scatter) reveals that the red wing of the C IV emission line is affected by strong intrinsic variability and microlensing (variability), as evidenced by the temporal changes in the wing.  The red wing of C III] is also subject to substantial intrinsic variability, while both the blue wings of C IV and C III] display moderate variations. In the blue wings of C IV and \mbox{C III]}, intrinsic variability affects both images to the same extent, while in the red wings, variability is slightly more pronounced in image B compared to image A, possibly due to microlensing variability. Mg II exhibits only limited signs of intrinsic variability and the effects of microlensing on this line are weak.

\begin{table}
\tabcolsep=0.36cm
\renewcommand{\arraystretch}{1.2}
\caption{Differential microlensing (B--A) in the BEL wings.}
\begin{tabular}{cccc} \hline \hline 
Emission Line & Wing & $\Delta$m (mag) & Scatter (mag) \\ 
(1) & (2) & (3) & (4) \\ \hline
\multirow{2}{*}{C IV} &  blue& $-0.07$ & $0.00$ \\
& red& $-0.33$ & $0.32$ \\ \hline 
\multirow{2}{*}{C III]} & blue & $\ \ 0.00$ & $0.14$ \\
& red & $+0.09$ & $0.33$ \\ \hline
\multirow{2}{*}{Mg II} & blue & -- & -- \\
& red & $-0.09$ & $0.07$ \\ \hline
\end{tabular} \\ 

\small \textbf{Notes.} Col. (1): Emission line core used as a reference for no microlensing. Col. (2): Emission line wing. Cols. (3)--(4): Average differential microlensing and scatter between observational epochs.
\label{BLR_ML}    
\end{table}

\begin{table}
\tabcolsep=0.17cm
\renewcommand{\arraystretch}{1.2}
\caption{Variability in the BEL wings across different observational epochs.}
\begin{tabular}{ccccc} \hline \hline 
Emission Line & Wing & Image & $\Delta$m (mag) & Scatter (mag) \\ 
(1) & (2) & (3) & (4) & (5) \\ \hline 
\multirow{4}{*}{C IV} & \multirow{2}{*}{blue} & A & $+0.06$ & $0.23$ \\
& & B & $+0.05$ & $0.30$ \\ \cline{2-5} 
& \multirow{2}{*}{red} & A & $-0.05$ & $0.72$\\ 
& & B & $-0.13$ & $0.66$ \\ \hline 
\multirow{4}{*}{C III]} & \multirow{2}{*}{blue} & A & $+0.02$ & $0.16$ \\
& & B & $+0.02$ & $0.18$ \\ \cline{2-5}
& \multirow{2}{*}{red} & A & $+0.02$ & $0.36$ \\
& & B & $+0.04$ & $0.47$ \\ \hline 
\multirow{4}{*}{Mg II} & \multirow{2}{*}{blue} & A & -- & -- \\
& & B & -- & -- \\ \cline{2-5} 
& \multirow{2}{*}{red} & A & $ \ \ 0.00$ & $0.09$\\
& & B & $-0.01$ &$ 0.11$ \\ \hline
\end{tabular}\\ 

\small \textbf{Notes.}  Col. (1): Emission line core used as a reference for no microlensing. Col. (2): Emission line wing. Col. (3): Lensed Image. Cols. (4)--(5): Average magnitude difference and scatter between epoch pairs.
\label{BLR_IV}    
\end{table}

\begin{figure*}[h]
\centering
\includegraphics[width=14cm]{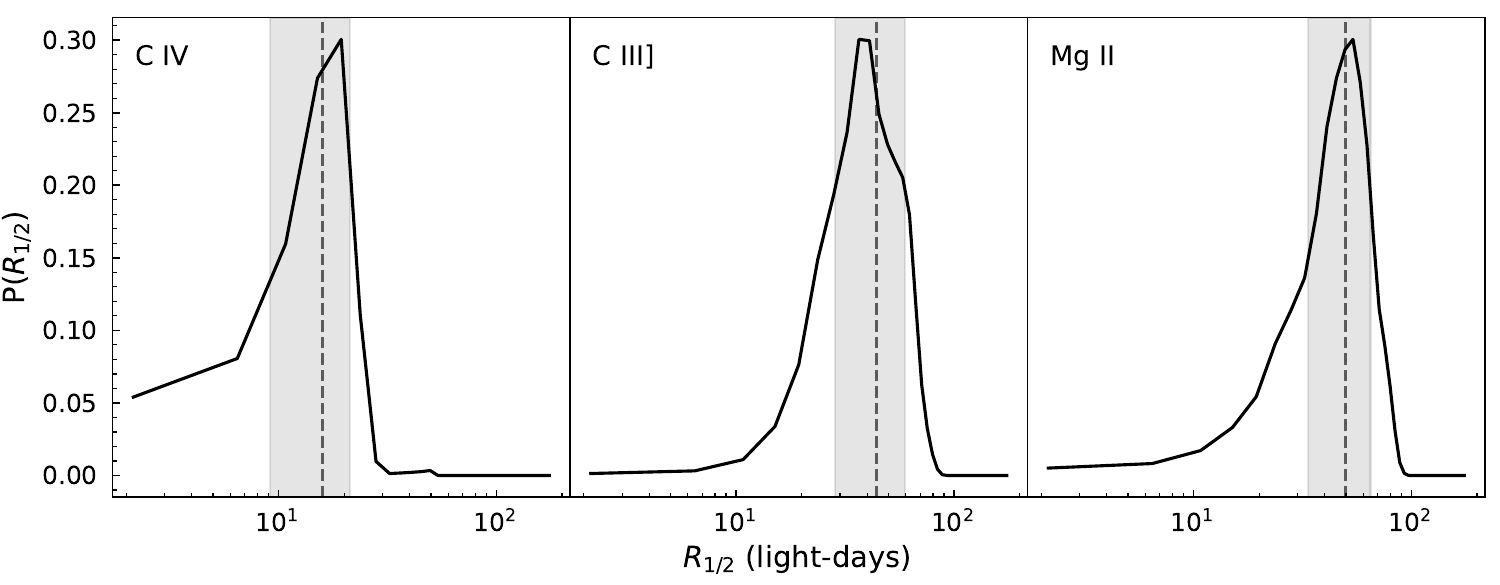}
\caption{Probability distributions of the half-light radii $R_{1/2}$ for the regions emitting the C IV (left), C III] (middle), and Mg II line (right). The vertical dashed lines indicate the expected size of the emission region, while the gray-shaded regions represent the one-sigma intervals.}
\label{PDFs}
\end{figure*}
\section{Bayesian size estimates}\label{4}
Based on the differential microlensing estimates between images A and B in the BEL wings, we can constrain the size of their emission region and provide insights into the BLR structure of the lensed quasar Q 0957+561. To obtain an average estimate of the size, we treat each microlensing measurement as a single-epoch event and compute the joint microlensing probability, $P(r_s)$, from all (time-delay-corrected and presumably independent) epochs of observation. Our procedure closely follows the methodology described in \citet{Guerras2013}.\\

Our simulations are based on 2000 $\times$ 2000 pixel$^2$ magnification maps, with a pixel size of approximately 0.3 light-days, covering a total area of 627$\times$627 light-days$^2$ on the source plane. The maps were generated using the Fast Multipole Method -- Inverse Polygon Mapping algorithm\footnote{https://gloton.ugr.es/microlensing/} (FMM--IPM) described by \citet{Jimenez2022}. The local convergence and shear values, $\kappa$ and $\gamma$, are used to determine the magnification at each image position. These dimensionless values were obtained by fitting a singular isothermal sphere with external shear (SIS+$\gamma_e$) to the image coordinates (\citealt{Mediavilla2009}).   
Although there are more sophisticated models available (see \citealt{Fadely2010}), this simple model was chosen to maintain consistency with prior publications on this source, including Paper I and II by \citet{Fian2021ad,Fian2022}. A mean stellar mass of $M = 0.3 M_{\odot}$ was assumed, and the size scale is proportional to the square root of the microlens mass ($r_s \propto \sqrt{M/M_\odot}$). The microlensing amplitude also depends on the local surface mass density fraction in compact objects (such as stars) compared to that of dark matter at the lensed image location (see, e.g., \citealt{Schechter2002}). In the case of Q 0957+561, the two lensed images are located at very different radii from the center of the lens galaxy, with image B appearing close to the center ($\sim$1\arcsec), and the light of component A passing far away ($\sim$5\arcsec). As a result, the surface mass density of the lens galaxy is much lower at the position of component A. In this study, we adopt the $\alpha$ values reported by \citet{Jimenez2015}, with $\alpha\sim0.3$ at the position of image B and $\alpha\sim0$ for image A, which is equivalent to not using any magnification map for this image. To investigate how variations in $\kappa$ and $\gamma$ affect the simulated magnification distribution of image B, we altered these parameters by $\pm0.1$. Our analysis reveals that varying $\kappa$ within this range does not result in significant changes to the microlensing distribution. When varying $\gamma$, we observe modest displacements of approximately $0.1$ mag.\\

To model the impact of extended sources on our simulations, we employ a Gaussian luminosity profile, $I(r_s) \propto exp(-R^2/2r_s^2)$, to represent the emission region of the BEL wings. We note that \citet{Mortonson2005} showed that the microlensing magnification statistics are largely insensitive to the radial profile of the source (see also \citealt{Munoz2016}). The magnifications experienced by a source of size $r_s$ are then obtained by convolving the magnification maps with 2D Gaussian profiles of sigma $r_s$. The probability of reproducing the observed microlensing is estimated by randomly placing the Gaussian source on our microlensing magnification maps. To cover a wide range of source sizes, we employ a linear grid ranging from $\sim$1 to 150 light-days. These sizes can be converted to half-light radii with the relation $R_{1/2} = 1.18\, r_s$.\\

Using our sample of spectroscopic measurements, we roughly determine the structure of the BLR in Q 0957+561. Adopting a Bayesian approach, we estimate the probability of $r_s$ given the observed microlensing magnifications. The resulting joint likelihood functions for the C IV, C III], and Mg II emission lines are displayed in Figure \ref{PDFs}, and allow us to calculate 68\% confidence size estimates. We infer sizes, expressed in $\sqrt{M/0.3M_\odot}$, of $R_{1/2} = 16.0_{-6.8}^{+5.3}$, $R_{1/2} = 44.3_{-16.0}^{+15.3}$, and $R_{1/2} = 49.9_{-16.4}^{+15.0}$ lt-days for the regions emitting the C IV, C III], and Mg II lines, respectively. To roughly examine the implications of covariance on our size estimates, we conducted an additional study for Mg II, excluding the epoch pair 5--8, which had less temporal separation ($\sim5$ years) than the other image pairs considered ($\sim$7 to $\sim$11 years). As a result, the uncertainties in the size estimate increases to $\pm18.3$ lt-days, reflecting a 17\% increase in the error compared to the initial measurement. \\

As mentioning above, the amplitude of microlensing is affected by the relative density of the local stellar surface mass compared to the dark matter at the image location, as reported by \citet{Schechter2002}. As a consequence, the emission region sizes are
also sensitive to the stellar mass fraction present, but the precise value of this latter ($\alpha \sim 0.3$) remains uncertain. To address this uncertainty, we studied the impact of $\alpha$ by varying it between 0.2 and 0.4. We recomputed the magnification maps and repeated the calculations, finding that the estimated sizes of the BEL emitting regions undergo a change of approximately 22\% on average. Specifically, we observe that larger values of $\alpha$ are associated with larger source sizes, whereas lower values of $\alpha$ lead to smaller size estimates. To explore the sensitivity of our size estimates to the width of the core region, we tested different intervals (1000, 2000, and 3000 km s$^{-1}$) to normalize the emission line cores. We then recalculated the sizes and evaluated the uncertainties introduced by this procedure. Our analysis shows that the size estimates vary by less than 10\%. We note that the separation of the line emission into two parts can be supported by the idea that the BLR is formed by a flat inner region (likely related to the accretion disk giving rise to the line wings), surrounded by a larger three-dimensional structure that produces the line core (see e.g., \citealt{Popovic2004}). Therefore, the microlensing-based size estimates for the regions emitting the line wings should be regarded as approximate lower bounds rather than a precise estimate of the size of the BEL region. It is important to mention that the blue wing of the semi-forbidden line \mbox{C III]} is contaminated by several underlying emission lines (Al III, Si III], and \mbox{Fe III}), which might bias the size estimates toward smaller sizes. Additionally, the high-velocity ends of the Mg II line wings may be affected by the pseudo-continuum formed by thousands of UV Fe II blends, and so caution should be taken when interpreting the results.
\section{Conclusions}\label{5}
Through optical spectroscopy, we conducted a search for microlensing and intrinsic variability in 15 epochs of observation of the gravitationally lensed quasar Q 0957+561. Our analysis of the flux ratios between images A and B in the three most frequently observed BELs provides insight into the inner structure of the quasar. Our main findings are as follows:

\begin{enumerate}
\item[1.] \textit{Extrinsic and intrinsic variability.} The techniques applied in this study enabled us to separately examine the effects of microlensing between the lensed quasar images A and B, and intrinsic variability within a single image across different observational epochs. Our findings indicate that \mbox{Q 0957+561} is a hybrid case with both moderate microlensing and strong intrinsic variability present. Consistent with prior studies (\citealt{Guerras2013,Fian2018blr,Fian2021blr}), our results show that high- and low-ionization lines are differently microlensed, with higher magnification seen in the high-ionization lines, indicating a more compact emission region. The measured microlensing in the red wing of C IV, averaging around 0.3 mag, is comparable to the microlensing (approximately 0.35 mag) determined from analyzing the 21year r-band light curves in this system (as per \citealt{Fian2021ad}). In general, intrinsic variability affects the same spectral features as microlensing (such as the red wing of C IV and to some extent the line wings of C III]), with varying levels of intensity depending on ionization degree. We find the impact of intrinsic variability or microlensing on the Mg II wings to be weak. 
\item[2.] \textit{BLR size.} Our analysis of the magnitude differences between the lensed images A and B suggests emission region sizes of $\gtrsim 44$ and $\gtrsim 50$ lt-days for C III] and Mg II, respectively, in line with previous studies (\citealt{Guerras2013,Fian2018blr}), as well as with findings from reverberation mapping campaigns (\citealt{Homayouni2020}). Our inferred size for the region emitting the C IV line wings ($\gtrsim 16.0$ lt-days) is comparable, within the errors, to that of the r-band continuum in this system ($\sim 17.6$ lt-days; see \citealt{Cornachione2020,Fian2021ad}), and to that of the UV Fe III blend ($\sim 15$ lt-days; see \citealt{Fian2022}), supporting the notion that the line partly originates from the accretion disk. By varying the stellar mass fraction, the lower limit of the C IV emission region changes to $14.0$ lt-days ($\alpha=0.2$) and $17.9$ lt-days ($\alpha=0.4$), respectively.
\item[3.] \textit{BLR geometry.} The impact of microlensing on the BLR of lensed quasars depends on the geometry and kinematics of the  BLR. Our spectroscopic analysis of the lensed quasar Q 0957+561 reveals microlensing effects on the broad-line components in the system. The extent of microlensing varies depending on the line, appearing in either the blue or red wing, or in both wings with equal intensity. This observation suggests that the BLR is not generally spherical in shape, in agreement with previous studies such as \citet{Sluse2012}, and recent findings by \citet{Fian2018blr, Fian2021blr} and \citet{Hutsemekers2021}. An anisotropic geometry or velocity field is required to produce asymmetrical deformations in an emission line, as demonstrated by \citet{Schneider1990}, \citet{Abajas2002}, and \citet{Lewis2004}. These latter authors showed that microlensing of a spherical BLR (in both geometry and velocity field) leads to symmetrical variations in the emission lines, while spatially separated approaching and receding parts of the velocity field in Keplerian disks can cause asymmetrical microlensing and possible shifts in the line centroid (as seen in \citealt{Braibant2016}). The asymmetry observed in the red wing of \mbox{C IV} might support the idea that this component originates from a compact region (a few light-days in size) with a nonspherical geometry, most likely following the motion of the accretion disk.\\
\end{enumerate}

\begin{acknowledgements} We thank the anonymous referee for the valuable comments and suggestions. We gratefully thank Luis J. Goicoechea and Vyacheslav N. Shalyapin for kindly providing most of the spectroscopic data listed in \mbox{Table \ref{data}}. This research was supported by the grants PID2020-118687GB-C31, PID2020-118687GB-C32, and PID2020-118687GB-C33, financed by the Spanish Ministerio de Ciencia e Innovación through MCIN/AEI/10.13039/501100011033. J.A.M. is also supported by the Generalitat Valenciana with the project of excellence Prometeo/2020/085. J.J.V. is also supported by projects FQM-108, P20\_00334, and A-FQM-510-UGR20/FEDER, financed by Junta de Andalucía.  V.M. acknowledges support from the project ANID FONDECYT Regular grant number 1231418, partial support from the Centro de Astrof\'{\i}sica de Valpara\'{\i}so, and from Fortalecimiento del Sistema de Investigaci\'on e Innovaci\'on de la Universidad de Valpara\'{\i}so (UVA20993). D.C. is financially supported by the DFG grant HA3555-14/1 to Tel Aviv University and University of Haifa, and by the Israeli Science Foundation grant no. 2398/19.
\end{acknowledgements}
\bibliographystyle{aa}
\bibliography{bibliography}

\begin{thebibliography}{53}
\expandafter\ifx\csname natexlab\endcsname\relax\def\natexlab#1{#1}\fi

\bibitem[{{Abajas} {et~al.}(2002){Abajas}, {Mediavilla}, {Mu{\~n}oz},
  {Popovi{\'c}}, \& {Oscoz}}]{Abajas2002}
{Abajas}, C., {Mediavilla}, E., {Mu{\~n}oz}, J.~A., {Popovi{\'c}}, L.~{\v{C}}.,
  \& {Oscoz}, A. 2002, \apj, 576, 640

\bibitem[{{Acebron} {et~al.}(2022){Acebron}, {Grillo}, {Bergamini}, {Caminha},
  {Tozzi}, {Mercurio}, {Rosati}, {Brammer}, {Meneghetti}, {Nonino}, \&
  {Vanzella}}]{Acebron2022}
{Acebron}, A., {Grillo}, C., {Bergamini}, P., {et~al.} 2022, \aap, 668, A142

\bibitem[{{Braibant} {et~al.}(2016){Braibant}, {Hutsem{\'e}kers}, {Sluse}, \&
  {Anguita}}]{Braibant2016}
{Braibant}, L., {Hutsem{\'e}kers}, D., {Sluse}, D., \& {Anguita}, T. 2016,
  \aap, 592, A23

\bibitem[{{Chang} \& {Refsdal}(1979)}]{Chang1979}
{Chang}, K. \& {Refsdal}, S. 1979, \nat, 282, 561

\bibitem[{{Cornachione} {et~al.}(2020{\natexlab{a}}){Cornachione}, {Morgan},
  {Burger}, {Shalyapin}, {Goicoechea}, {Vrba}, {Dahm}, \&
  {Tilleman}}]{Cornachione2020}
{Cornachione}, M.~A., {Morgan}, C.~W., {Burger}, H.~R., {et~al.}
  2020{\natexlab{a}}, \apj, 905, 7

\bibitem[{{Cornachione} {et~al.}(2020{\natexlab{b}}){Cornachione}, {Morgan},
  {Millon}, {Bentz}, {Courbin}, {Bonvin}, \& {Falco}}]{Cornachione2020b}
{Cornachione}, M.~A., {Morgan}, C.~W., {Millon}, M., {et~al.}
  2020{\natexlab{b}}, \apj, 895, 125

\bibitem[{{Fadely} {et~al.}(2010){Fadely}, {Keeton}, {Nakajima}, \&
  {Bernstein}}]{Fadely2010}
{Fadely}, R., {Keeton}, C.~R., {Nakajima}, R., \& {Bernstein}, G.~M. 2010,
  \apj, 711, 246

\bibitem[{{Fian} {et~al.}(2018{\natexlab{a}}){Fian}, {Guerras}, {Mediavilla},
  {Jim{\'e}nez-Vicente}, {Mu{\~n}oz}, {Falco}, {Motta}, \&
  {Hanslmeier}}]{Fian2018blr}
{Fian}, C., {Guerras}, E., {Mediavilla}, E., {et~al.} 2018{\natexlab{a}}, \apj,
  859, 50

\bibitem[{{Fian} {et~al.}(2016){Fian}, {Mediavilla}, {Hanslmeier}, {Oscoz},
  {Serra-Ricart}, {Mu{\~n}oz}, \& {Jim{\'e}nez-Vicente}}]{Fian2016}
{Fian}, C., {Mediavilla}, E., {Hanslmeier}, A., {et~al.} 2016, \apj, 830, 149

\bibitem[{{Fian} {et~al.}(2021{\natexlab{a}}){Fian}, {Mediavilla},
  {Jim{\'e}nez-Vicente}, {Motta}, {Mu{\~n}oz}, {Chelouche},
  {Gom{\'e}z-Alvarez}, {Rojas}, \& {Hanslmeier}}]{Fian2021ad}
{Fian}, C., {Mediavilla}, E., {Jim{\'e}nez-Vicente}, J., {et~al.}
  2021{\natexlab{a}}, \aap, 654, A70

\bibitem[{{Fian} {et~al.}(2022){Fian}, {Mediavilla}, {Jim{\'e}nez-Vicente},
  {Motta}, {Mu{\~n}oz}, {Chelouche}, \& {Hanslmeier}}]{Fian2022}
{Fian}, C., {Mediavilla}, E., {Jim{\'e}nez-Vicente}, J., {et~al.} 2022, \aap,
  667, A67

\bibitem[{{Fian} {et~al.}(2018{\natexlab{b}}){Fian}, {Mediavilla},
  {Jim{\'e}nez-Vicente}, {Mu{\~n}oz}, \& {Hanslmeier}}]{Fian2018ad}
{Fian}, C., {Mediavilla}, E., {Jim{\'e}nez-Vicente}, J., {Mu{\~n}oz}, J.~A., \&
  {Hanslmeier}, A. 2018{\natexlab{b}}, \apj, 869, 132

\bibitem[{{Fian} {et~al.}(2021{\natexlab{b}}){Fian}, {Mediavilla}, {Motta},
  {Jim{\'e}nez-Vicente}, {Mu{\~n}oz}, {Chelouche}, \&
  {Hanslmeier}}]{Fian2021blr}
{Fian}, C., {Mediavilla}, E., {Motta}, V., {et~al.} 2021{\natexlab{b}}, \aap,
  653, A109

\bibitem[{{For{\'e}s-Toribio} {et~al.}(2022){For{\'e}s-Toribio}, {Mu{\~n}oz},
  {Kochanek}, \& {Mediavilla}}]{Fores2022}
{For{\'e}s-Toribio}, R., {Mu{\~n}oz}, J.~A., {Kochanek}, C.~S., \&
  {Mediavilla}, E. 2022, \apj, 937, 35

\bibitem[{{Garrett} {et~al.}(1992){Garrett}, {Walsh}, \&
  {Carswell}}]{Garrett1992}
{Garrett}, M.~A., {Walsh}, D., \& {Carswell}, R.~F. 1992, \mnras, 254, 27P

\bibitem[{{Gil-Merino} {et~al.}(2018){Gil-Merino}, {Goicoechea}, {Shalyapin},
  \& {Oscoz}}]{GilMerino2018}
{Gil-Merino}, R., {Goicoechea}, L.~J., {Shalyapin}, V.~N., \& {Oscoz}, A. 2018,
  \aap, 616, A118

\bibitem[{{Goicoechea} {et~al.}(2005){Goicoechea}, {Gil-Merino}, \&
  {Ull{\'a}n}}]{Goicoechea2005}
{Goicoechea}, L.~J., {Gil-Merino}, R., \& {Ull{\'a}n}, A. 2005, \mnras, 360,
  L60

\bibitem[{{Gondhalekar} \& {Wilson}(1982)}]{Gondhalekar1982}
{Gondhalekar}, P.~M. \& {Wilson}, R. 1982, \nat, 296, 415

\bibitem[{{Guerras} {et~al.}(2013){Guerras}, {Mediavilla}, {Jimenez-Vicente},
  {Kochanek}, {Mu{\~n}oz}, {Falco}, \& {Motta}}]{Guerras2013}
{Guerras}, E., {Mediavilla}, E., {Jimenez-Vicente}, J., {et~al.} 2013, \apj,
  764, 160

\bibitem[{{Homayouni} {et~al.}(2020){Homayouni}, {Trump}, {Grier}, {Horne},
  {Shen}, {Brandt}, {Dawson}, {Alvarez}, {Green}, {Hall}, {Hern{\'a}ndez
  Santisteban}, {Ho}, {Kinemuchi}, {Kochanek}, {Li}, {Peterson}, {Schneider},
  {Starkey}, {Bizyaev}, {Pan}, {Oravetz}, \& {Simmons}}]{Homayouni2020}
{Homayouni}, Y., {Trump}, J.~R., {Grier}, C.~J., {et~al.} 2020, \apj, 901, 55

\bibitem[{{Hutchings}(2003)}]{Hutchings2003}
{Hutchings}, J.~B. 2003, \aj, 126, 24

\bibitem[{{Hutsem{\'e}kers} \& {Sluse}(2021)}]{Hutsemekers2021}
{Hutsem{\'e}kers}, D. \& {Sluse}, D. 2021, \aap, 654, A155

\bibitem[{{Jim{\'e}nez-Vicente} \& {Mediavilla}(2022)}]{Jimenez2022}
{Jim{\'e}nez-Vicente}, J. \& {Mediavilla}, E. 2022, \apj, 941, 80

\bibitem[{{Jim{\'e}nez-Vicente} {et~al.}(2015){Jim{\'e}nez-Vicente},
  {Mediavilla}, {Kochanek}, \& {Mu{\~n}oz}}]{Jimenez2015}
{Jim{\'e}nez-Vicente}, J., {Mediavilla}, E., {Kochanek}, C.~S., \& {Mu{\~n}oz},
  J.~A. 2015, \apj, 799, 149

\bibitem[{{Lewis} \& {Ibata}(2004)}]{Lewis2004}
{Lewis}, G.~F. \& {Ibata}, R.~A. 2004, \mnras, 348, 24

\bibitem[{{Liao}(2020)}]{Liao2020}
{Liao}, K. 2020, \apjl, 899, L33

\bibitem[{{Liu} {et~al.}(2022){Liu}, {Cao}, {Li}, {Zheng}, {Liu}, {Guo}, \&
  {Zheng}}]{Liu2022}
{Liu}, T., {Cao}, S., {Li}, X., {et~al.} 2022, \aap, 668, A51

\bibitem[{{Lloyd}(1981)}]{Lloyd1981}
{Lloyd}, C. 1981, \nat, 294, 727

\bibitem[{{Mediavilla} {et~al.}(2018){Mediavilla}, {Jim{\'e}nez-Vicente},
  {Fian}, {Mu{\~n}oz}, {Falco}, {Motta}, \& {Guerras}}]{Mediavilla2018}
{Mediavilla}, E., {Jim{\'e}nez-Vicente}, J., {Fian}, C., {et~al.} 2018, \apj,
  862, 104

\bibitem[{{Mediavilla} {et~al.}(2019){Mediavilla}, {Jim{\'e}nez-vicente},
  {Mej{\'\i}a-restrepo}, {Motta}, {Falco}, {Mu{\~n}oz}, {Fian}, \&
  {Guerras}}]{Mediavilla2019}
{Mediavilla}, E., {Jim{\'e}nez-vicente}, J., {Mej{\'\i}a-restrepo}, J.,
  {et~al.} 2019, \apj, 880, 96

\bibitem[{{Mediavilla} {et~al.}(2016){Mediavilla}, {Jim{\'e}nez-Vicente},
  {Mu{\~n}oz}, \& {Battaner}}]{Mediavilla2016}
{Mediavilla}, E., {Jim{\'e}nez-Vicente}, J., {Mu{\~n}oz}, J.~A., \& {Battaner},
  E. 2016, \apj, 832, 46

\bibitem[{{Mediavilla} {et~al.}(2009){Mediavilla}, {Mu{\~n}oz}, {Falco},
  {Motta}, {Guerras}, {Canovas}, {Jean}, {Oscoz}, \&
  {Mosquera}}]{Mediavilla2009}
{Mediavilla}, E., {Mu{\~n}oz}, J.~A., {Falco}, E., {et~al.} 2009, \apj, 706,
  1451

\bibitem[{{Miller} {et~al.}(1981){Miller}, {Antonucci}, \& {Keel}}]{Miller1981}
{Miller}, J.~S., {Antonucci}, R.~R.~J., \& {Keel}, W.~C. 1981, \nat, 289, 153

\bibitem[{{Millon} {et~al.}(2020){Millon}, {Galan}, {Courbin}, {Treu}, {Suyu},
  {Ding}, {Birrer}, {Chen}, {Shajib}, {Sluse}, {Wong}, {Agnello}, {Auger},
  {Buckley-Geer}, {Chan}, {Collett}, {Fassnacht}, {Hilbert}, {Koopmans},
  {Motta}, {Mukherjee}, {Rusu}, {Sonnenfeld}, {Spiniello}, \& {Van de
  Vyvere}}]{Millon2020}
{Millon}, M., {Galan}, A., {Courbin}, F., {et~al.} 2020, \aap, 639, A101

\bibitem[{{Mortonson} {et~al.}(2005){Mortonson}, {Schechter}, \&
  {Wambsganss}}]{Mortonson2005}
{Mortonson}, M.~J., {Schechter}, P.~L., \& {Wambsganss}, J. 2005, \apj, 628,
  594

\bibitem[{{Mosquera} {et~al.}(2011){Mosquera}, {Mu{\~n}oz}, {Mediavilla}, \&
  {Kochanek}}]{Mosquera2011}
{Mosquera}, A.~M., {Mu{\~n}oz}, J.~A., {Mediavilla}, E., \& {Kochanek}, C.~S.
  2011, \apj, 728, 145

\bibitem[{{Motta} {et~al.}(2012){Motta}, {Mediavilla}, {Falco}, \&
  {Mu{\~n}oz}}]{Motta2012}
{Motta}, V., {Mediavilla}, E., {Falco}, E., \& {Mu{\~n}oz}, J.~A. 2012, \apj,
  755, 82

\bibitem[{{Mu{\~{n}}oz} {et~al.}(2016){Mu{\~{n}}oz}, {Vives-Arias}, {Mosquera},
  {Jim{\'e}nez-Vicente}, {Kochanek}, \& {Mediavilla}}]{Munoz2016}
{Mu{\~{n}}oz}, J.~A., {Vives-Arias}, H., {Mosquera}, A.~M., {et~al.} 2016,
  \apj, 817, 155

\bibitem[{{Napier} {et~al.}(2023){Napier}, {Sharon}, {Dahle}, {Bayliss},
  {Gladders}, {Mahler}, {Rigby}, \& {Florian}}]{Napier2023}
{Napier}, K., {Sharon}, K., {Dahle}, H., {et~al.} 2023, arXiv e-prints,
  arXiv:2301.11240

\bibitem[{{Paczynski}(1986)}]{Paczynski1986}
{Paczynski}, B. 1986, \apj, 304, 1

\bibitem[{{Planesas} {et~al.}(1999){Planesas}, {Martin-Pintado}, {Neri}, \&
  {Colina}}]{Planesas1999}
{Planesas}, P., {Martin-Pintado}, J., {Neri}, R., \& {Colina}, L. 1999,
  Science, 286, 2493

\bibitem[{{Popovi{\'c}} {et~al.}(2004){Popovi{\'c}}, {Mediavilla}, {Bon}, \&
  {Ili{\'c}}}]{Popovic2004}
{Popovi{\'c}}, L.~{\v{C}}., {Mediavilla}, E., {Bon}, E., \& {Ili{\'c}}, D.
  2004, \aap, 423, 909

\bibitem[{{Rojas} {et~al.}(2020){Rojas}, {Motta}, {Mediavilla},
  {Jim{\'e}nez-Vicente}, {Falco}, \& {Fian}}]{Rojas2020}
{Rojas}, K., {Motta}, V., {Mediavilla}, E., {et~al.} 2020, \apj, 890, 3

\bibitem[{{Schechter} \& {Wambsganss}(2002)}]{Schechter2002}
{Schechter}, P.~L. \& {Wambsganss}, J. 2002, \apj, 580, 685

\bibitem[{{Schneider} \& {Wambsganss}(1990)}]{Schneider1990}
{Schneider}, P. \& {Wambsganss}, J. 1990, \aap, 237, 42

\bibitem[{{Shalyapin} {et~al.}(2008){Shalyapin}, {Goicoechea}, {Koptelova},
  {Ull{\'a}n}, \& {Gil-Merino}}]{Shalyapin2008}
{Shalyapin}, V.~N., {Goicoechea}, L.~J., {Koptelova}, E., {Ull{\'a}n}, A., \&
  {Gil-Merino}, R. 2008, \aap, 492, 401

\bibitem[{{Sluse} {et~al.}(2012){Sluse}, {Hutsem{\'e}kers}, {Courbin},
  {Meylan}, \& {Wambsganss}}]{Sluse2012}
{Sluse}, D., {Hutsem{\'e}kers}, D., {Courbin}, F., {Meylan}, G., \&
  {Wambsganss}, J. 2012, \aap, 544, A62

\bibitem[{{Stockton}(1980)}]{Stockton1980}
{Stockton}, A. 1980, \apjl, 242, L141

\bibitem[{{Tie} \& {Kochanek}(2018)}]{Tie2018}
{Tie}, S.~S. \& {Kochanek}, C.~S. 2018, \mnras, 473, 80

\bibitem[{{Walsh} {et~al.}(1979){Walsh}, {Carswell}, \& {Weymann}}]{Walsh1979}
{Walsh}, D., {Carswell}, R.~F., \& {Weymann}, R.~J. 1979, \nat, 279, 381

\bibitem[{{Wang} {et~al.}(2022){Wang}, {Zhang}, {He}, {Zhang}, \&
  {Zhang}}]{Wang2022}
{Wang}, L.-F., {Zhang}, J.-H., {He}, D.-Z., {Zhang}, J.-F., \& {Zhang}, X.
  2022, \mnras, 514, 1433

\bibitem[{{Wong} {et~al.}(2020){Wong}, {Suyu}, {Chen}, {Rusu}, {Millon},
  {Sluse}, {Bonvin}, {Fassnacht}, {Taubenberger}, {Auger}, {Birrer}, {Chan},
  {Courbin}, {Hilbert}, {Tihhonova}, {Treu}, {Agnello}, {Ding}, {Jee},
  {Komatsu}, {Shajib}, {Sonnenfeld}, {Bland ford}, {Koopmans}, {Marshall}, \&
  {Meylan}}]{Wong2020}
{Wong}, K.~C., {Suyu}, S.~H., {Chen}, G. C.~F., {et~al.} 2020, \mnras
  [\eprint[arXiv]{1907.04869}]

\bibitem[{{Zuo} {et~al.}(1997){Zuo}, {Beaver}, {Burbidge}, {Cohen},
  {Junkkarinen}, \& {Lyons}}]{Zuo1997}
{Zuo}, L., {Beaver}, E.~A., {Burbidge}, E.~M., {et~al.} 1997, \apj, 477, 568

\end{thebibliography}

\end{document}